\documentclass[10pt,aps,prb,twocolumn, nofootinbib]{revtex4-1}
\usepackage{amssymb,amsmath}
\usepackage{subfig}
\usepackage{graphicx}
\usepackage[font=small,labelfont=bf,
   justification=justified,
   format=plain]{caption}
\usepackage{color}
\usepackage{hyperref}
\usepackage{listings}
\hypersetup{
    bookmarks=true,         
    unicode=false,          
    pdftoolbar=true,        
    pdfmenubar=true,        
    pdffitwindow=false,     
    pdfstartview={FitH},    
    pdfauthor={Kevin Multani},     
    colorlinks=true,       
    linkcolor=blue,          
    citecolor=red,        
}
\graphicspath{{figures/}}

\begin{document}

\definecolor{dkgreen}{rgb}{0,0.6,0}
\definecolor{gray}{rgb}{0.5,0.5,0.5}
\definecolor{mauve}{rgb}{0.58,0,0.82}
\bibliographystyle{unsrt}
\captionsetup{justification=raggedright,singlelinecheck=false}
\lstset{frame=tb,
  	language=Matlab,
  	aboveskip=3mm,
  	belowskip=3mm,
  	showstringspaces=false,
  	columns=flexible,
  	basicstyle={\small\ttfamily},
  	numbers=none,
  	numberstyle=\tiny\color{gray},
 	keywordstyle=\color{blue},
	commentstyle=\color{dkgreen},
  	stringstyle=\color{mauve},
  	breaklines=true,
  	breakatwhitespace=true
  	tabsize=3
}

\title{Bound States of Charged Adatoms on MoS$_2$: Screening and Multivalley Effects}

\author{Martik Aghajanian, Arash A. Mostofi, Johannes Lischner}
\affiliation{Depts. of Physics and Materials and the Thomas Young Centre for Theory and Simulation of Materials, Imperial College London, London, SW7 2AZ, UK}
\date{\today}

\begin{abstract}
Adsorbate engineering is a promising route for controlling the electronic properties of monolayer transition-metal dichalcogenide materials. Here, we study shallow bound states induced by charged adatoms on MoS$_2$ using large-scale tight-binding simulations with screened adatom potentials obtained from ab initio calculations. The interplay of unconventional screening in two-dimensional systems and multivalley effects in the transition-metal dichalcogenide (TMDC) band structure results in a rich diversity of bound impurity states. We present results for impurity state wavefunctions and energies, as well as for the local density of states in the vicinity of the adatom which can be measured using scanning tunnelling spectroscopy. We find that the presence of several distinct valleys in the MoS$_2$ band structure gives rise to crossovers of impurity states at critical charge strengths, altering the orbital character of the most strongly bound state. We compare our results to simpler methods, such as the 2D hydrogen atom and effective mass theory, and we discuss limitations of these approaches.
\end{abstract}
\maketitle

Since the discovery of graphene, there has been significant interest in the development of ultrathin devices based on two-dimensional (2D) materials. In contrast to graphene, which is a semimetal when undoped, monolayer transition-metal dichalcogenides (TMDCs) with the chemical formula $MX_2$ ($M$=Mo, W; $X$=S, Se, Te) are semiconductors with a direct band gap \cite{liu15,Kadantsev2012}. Monolayer TMDCs have been used as channel materials in field-effect transistors\cite{Wang2012,Xu2017} and microprocessors\cite{Wachter2017}, as well as absorbers in solar cells\cite{Baugher2014} and as sensors\cite{LopezSanchez2013,Kalantar2016}, with promising results.

Defects play a critical role in the performance of devices under realistic conditions \cite{Review2015,kuc15,SchmidtMende2007}. Analogously to conventional bulk semiconductors, impurities with shallow donor or acceptor states can be used to control the carrier concentration in TMDCs via defect engineering\cite{Queisser1998,Janotti2009}. Adsorbed atoms and molecules are a particularly promising class of impurities in TMDCs as they tend to only weakly perturb the atomic structure of the TMDC substrate, thereby limiting any degradation of carrier mobility that may result from impurity scattering or trapping\cite{Chin2012,Leijtens2016}, and experimental fabrication of adsorbate-engineered samples is straightforward\cite{Komesu2017}. 

A detailed theoretical understanding of the properties of adsorbates on TMDCs is important to enable the rational design of new devices. On the one hand, many groups have used ab initio density-functional theory (DFT) to study the interaction of adsorbed atoms and molecules with TMDCs. Such calculations yield important material-specific insights about adsorption geometries, adsorbate binding energies and charge transfer \cite{Rastogi2014,chang14,Ataca2011,Fang2013,Douloui2013}. However, ab initio calculations are limited in terms of the size of the systems that can be considered (typically containing up to several hundred or a few thousand atoms), which are much too small to describe properties of shallow defect states that can extend up 100 \AA ngstrom (\AA) or more, as has been observed recently for Coulomb impurities in graphene using scanning tunnelling spectroscopy (STS) \cite{Wong2017}. 

On the other hand, continuum electronic structure methods, such as Dirac theory for graphene or effective mass theory for bulk semiconductors, can describe the behaviour of extended impurity states, but require parameters from experiments or ab initio calculations, such as Fermi velocities, effective masses \cite{corsetti17,Bassani1969,kohn1957,Rak2009,Shimizu1965} and rather importantly, the defect potential that is typically screened by electrons of the host material. 

In this paper, we study properties of shallow impurity states induced by charged adatoms on monolayer MoS$_2$. Using large-scale tight-binding models and screened defect potentials calculated from ab initio dielectric functions, we reveal a surprising diversity of bound defect states resulting from the unconventional screening present in reduced-dimensional materials and the interplay between multiple valleys in the TMDC band structure. We present results for impurity wavefunctions and binding energies as function of the impurity charge and also compute the local density of states (LDOS) in the vicinity of the adatom, which can be measured in STS experiments. For both donor and acceptor impurities, we find that impurity wavefunctions have similar nodal structure to 2D hydrogenic states, but with radii that lie on the nanoscale. We find that that the orbital character of the most strongly bound impurity state switches as a function of the impurity charge strength $Z$ due to the different effective masses associated with different valleys in the monolayer TMDC band structure.  We compare our results to the 2D hydrogen atom and also to effective mass theory calculations and discuss the limitations of these continuum models. Whilst an approach based on the effective mass model is able to describe some of the general behaviour with reasonable accuracy, we find significant discrepancies from our tight-binding model which arise from short-range features of the defect potential. Our calculations demonstrate the potential of adsorbate engineering for ultrathin devices based on TMDCs and the importance of first-principles based description of their properties.

\textbf{Modelling charged adatoms on MoS$_2$} - To describe the electronic structure of the MoS$_2$ monolayer, we employ the three-band tight-binding (TB) model by Liu \emph{et al.} \cite{liu13}. This model uses a basis of transition-metal $4d_{z^2}$, $4d_{xy}$ and $4d_{x^2-y^2}$ orbitals which give the dominant contribution to the states near the conduction and valence band extrema and includes hoppings up to third-nearest neighbours as well as spin-orbit interactions. The various parameters were determined by fits to DFT band structures. 

\begin{figure}[t!]
\includegraphics[]{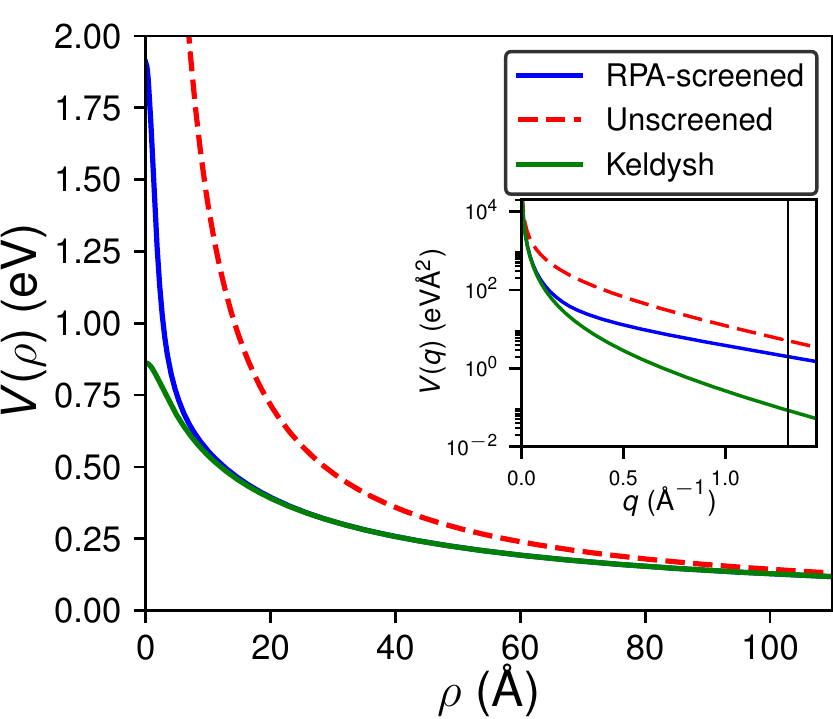}
\caption{RPA-screened potential of a charged adatom situated $d=2\text{ \AA }$ above the Mo-atom in MoS$_2$ with strength $Z=1$ (blue solid curve) compared to the unscreened Coulomb potential (red dashed curve). We also compare this to the Keldysh model (green curve) for $Z=1$, $d=2\text{ \AA}$ and screening length $\rho_0=45\text{ \AA}$ (Eq.~\ref{eqn:screen}), fitted to the RPA-screened potential. The inset shows the Fourier transform of the screened and unscreened potentials, as well as the potential screened in the Keldysh model, with the solid vertical line indicating $|\mathbf{K}-\mathbf{K}^{\prime}|$, the separation in reciprocal space between the two valleys of MoS$_2$.}
\label{fig:eps}
\end{figure}

The charged adatom is described as a point charge $Q=Ze$ (with $e$ being the proton charge) located a distance $d$ above the plane of the transition-metal atoms. The charge gives rise to a screened potential in the TMDC sheet. Within linear response theory, the screened potential is given by 
\begin{equation}\label{eqn:hankel}
V(\rho;\,Z, d) = Ze^2\int_0^{\infty}\mathrm{d}q\; \varepsilon^{-1}_{\text{2D}}(q)J_0(q \rho )\,e^{-qd},
\end{equation}
where $\rho$ denotes the in-plane distance from the adatom and $\varepsilon^{-1}_{\text{2D}}(q)$ is the inverse 2D dielectric function of a single TMDC monolayer. The 2D dielectric function can be obtained from the inverse dielectric matrix $\varepsilon^{-1}_{\mathbf{G} \mathbf{G}^{\prime}}(\mathbf{q})$ of an infinite system of stacked TMDC sheets (simulated in an electronic structure calculation that employs periodic boundary conditions) via \cite{Qui2016}
\begin{equation}
\varepsilon^{-1}_{\text{2D}}(\mathbf{q}) = \frac{q}{2\pi e^2 L_z}\sum_{\mathbf{G}_z \mathbf{G}^{\prime}_z}\varepsilon^{-1}_{\mathbf{G}_z \mathbf{G}^{\prime}_z}(\mathbf{q})v_{\text{trunc}}(|\mathbf{q}+\mathbf{G}^{\prime}_z|).\label{eqn:eps2d}
\end{equation}
Here, $\mathbf{G}_z$ and $\mathbf{G}^{\prime}_z$ denote reciprocal lattice vectors along the out-of-plane ($z$) direction, $v_{\text{trunc}}$ is a slab-truncated Coulomb interaction\cite{Ismail2006} and $L_z$ denotes the distance between the stacked sheets. The inverse dielectric matrix is computed for a MoS$_2$ monolayer using the random-phase approximation\cite{BohmPines} (RPA) with Kohn-Sham wave functions and energies from ab initio DFT (see supplementary materials for details). Calculations were carried out using the Quantum Espresso\cite{Giannozzi2009} and BerkeleyGW software packages \cite{deslippe2012}. For small wave vectors, which are relevant for describing shallow impurity impurity bound states, we find that the right hand side of Eq.~\eqref{eqn:eps2d} depends only on the \emph{magnitude} of the wave vector. Fig.~\ref{fig:eps} shows the screened (calculated from Eq.\ref{eqn:hankel}) and unscreened potentials of a charged adatom with $Z=1$ and $d=2\text{ \AA }$ above the Mo-layer in the MoS$_2$ sheet. While there are clear differences at short distances, the two potentials both converge to the unscreened case at long distances from the adatom which is characteristic of screening in 2D semiconductors. This short-range discrepancy corresponds to significant differences between the Fourier transforms of these potentials at large wavevectors, shown in the inset of Fig~\ref{fig:eps}.

To study shallow bound states of the screened adatom potential, we construct a $51 \times 51$ TMDC supercell containing $7803$ atoms and diagonalize the resulting TB Hamiltonian with the adatom potential as an on-site term\cite{corsetti17,Wong2017}. Note that the adatom is placed above a transition-metal site as this is the preferred adsorption geometry for many adatom species, such as alkali metals\cite{Rastogi2014,chang14,Ataca2011}.

To analyze the results of our atomistic tight-binding simulations, we have also carried out calculations using effective mass theory. In this approach, which has been used routinely to study shallow bound states of charged impurities in bulk semiconductors \cite{Lanoo92,LiXia2007,kohn1957}, the impurity states are expressed as $\Psi_{n\nu}(\mathbf{r}) = \int\mathrm{d}\mathbf{k}\;\phi_{n\nu}(\mathbf{k})\psi_{n\mathbf{k}}(\mathbf{r})$. Here, $\psi_{n\mathbf{k}}$ denotes an unperturbed Bloch state with band index $n$ and crystal momentum $\mathbf{k}$ of the host material and $\phi_{n\nu}(\mathbf{k})$ is an envelope function determined by \cite{Bassani1974}
\begin{align}
\epsilon_{n\mathbf{k}}\phi_{n\nu}(\mathbf{k}) + \int\mathrm{d}\mathbf{k'}\;\langle\psi_{n\mathbf{k}}|V|\psi_{n\mathbf{k'}}\rangle \phi_{n\nu}(\mathbf{k'})=E_{n\nu} \phi_{n\nu}(\mathbf{k}),
\label{eqn:blochimp}
\end{align}
where $\epsilon_{n\mathbf{k}}$ describes the band structure of the host material and $V(\mathbf{r})$ denotes the screened impurity potential. In bulk semiconductors, $V$ can be accurately approximated\cite{Bassani1974} by $Ze^2\varepsilon^{-1}(q=0)/r$ and the resulting equation for the impurity state envelope function reduces to the Schr\"odinger equation of a hydrogen atom with a reduced Bohr radius $\widetilde{a}_0=(m^*/m_0)Za_{0}\varepsilon^{-1}(q=0)$ (with $m^*$ and $m_0$ denoting the effective and bare mass of the electron, respectively, and $a_0$ is the Bohr radius). In this approximation, the impurity state envelope functions take the form of the 2D hydrogenic states\cite{Yang1991} give by
\begin{equation}
\phi_{nl}^{(\text{2DH})}(\rho, \theta) = \frac{e^{il\theta}}{N_{nl}(Z, m^*)}\left(\rho\lambda_n\right)^{|l|}e^{\rho\lambda_n/2}L_{n-l-1}^{2|l|}\left(\rho\lambda_n\right),
\end{equation}
where $N_{nl}$ is a normalization constant, $L_j^k$ are the generalized Laguerre polynomials, and $\lambda_n=\frac{2}{2n+1}\frac{Zm^*e^2}{4\pi\varepsilon_0\hbar^2}$. We compare these solutions to the wavefunctions extracted from our TB model to identify similarities in nodal structure.

The screened impurity potential in a 2D semiconductor, such as a TMDC monolayer, however, cannot be accurately approximated by a bare Coulomb interaction divided by a \emph{constant} dielectric function (see Fig.~\ref{fig:eps}). A well-known model for the screening of a point charge embedded in a thin dielectric film was derived by Keldysh\cite{Keldysh79} and is given by

\begin{equation}\label{eqn:screen}
\varepsilon_{\text{Keldysh}}(q) = 1+\rho_0 q
\end{equation}
where $\rho_0$ is the screening length. We calculate the screened potential $V_{\text{Keldysh}}(\rho)$ using the Keldysh model by substituting $\varepsilon^{-1}_{\text{Keldysh}}(q)$ for the inverse dielectric function in Eq.~\ref{eqn:hankel}. The value of $\rho_0=45\text{ \AA }$ is obtained by fitting to the RPA-screened potential of Fig~\ref{fig:eps}. The Keldysh model has been frequently used to study excitons in TMDCs\cite{Berkelbach2013,Qui2016,Cudazzo2010} and we also use it here for comparison to our tight-binding results.

To simplify the integration over $k$-points in Eq.~\eqref{eqn:blochimp}, Bassani \emph{et al.}\cite{Bassani1974} divided the first Brillouin zone into subzones $\Omega_i$ centered on critial points $\mathbf{k}_i$, typically associated with band extrema. The impurity states $\Psi_{n\nu}(\mathbf{r})$ are then constructed as linear combinations of subzone states 
\begin{equation}
\Psi_{n\nu i}(\mathbf{r})\approx \phi_{n\nu i}(\mathbf{r})\psi_{n\mathbf{k}_i}(\mathbf{r})\label{eqn:Ryd}.
\end{equation}
To determine the subzone envelope functions $\phi_{n\nu i}(\mathbf{r})$, we minimize the expectation value of the Keldysh Hamiltonian $\hat{H} =\frac{-\hbar^2}{2m_i^{\ast}}(\partial^2_x+\partial_y^2) + V_{\text{Keldysh}}(r)$ (where $m^*_i$ denotes the effective mass associated with the relevant conduction or valence band at $\mathbf{k}_i$) using the following ansatz for the most strongly bound impurity state
\begin{equation}
\phi_{1s,i}(\rho; \alpha) = \frac{(2\alpha)e^{\alpha d}}{\sqrt{2\pi(2\alpha d + 1)}}e^{-\alpha\sqrt{\rho^2+d^2}},
\label{eqn:envelope}
\end{equation}
where $\alpha$ is a variational parameter, which we use to define the impurity radius $a_{\text{imp}}=\alpha^{-1}$. Once the subzone states are obtained, the full impurity states are found by including interactions between different subzones. As the coupling is usually weak, it can be treated using perturbation theory \cite{Bassani1974}.

\begin{figure}[ht!]
\includegraphics[]{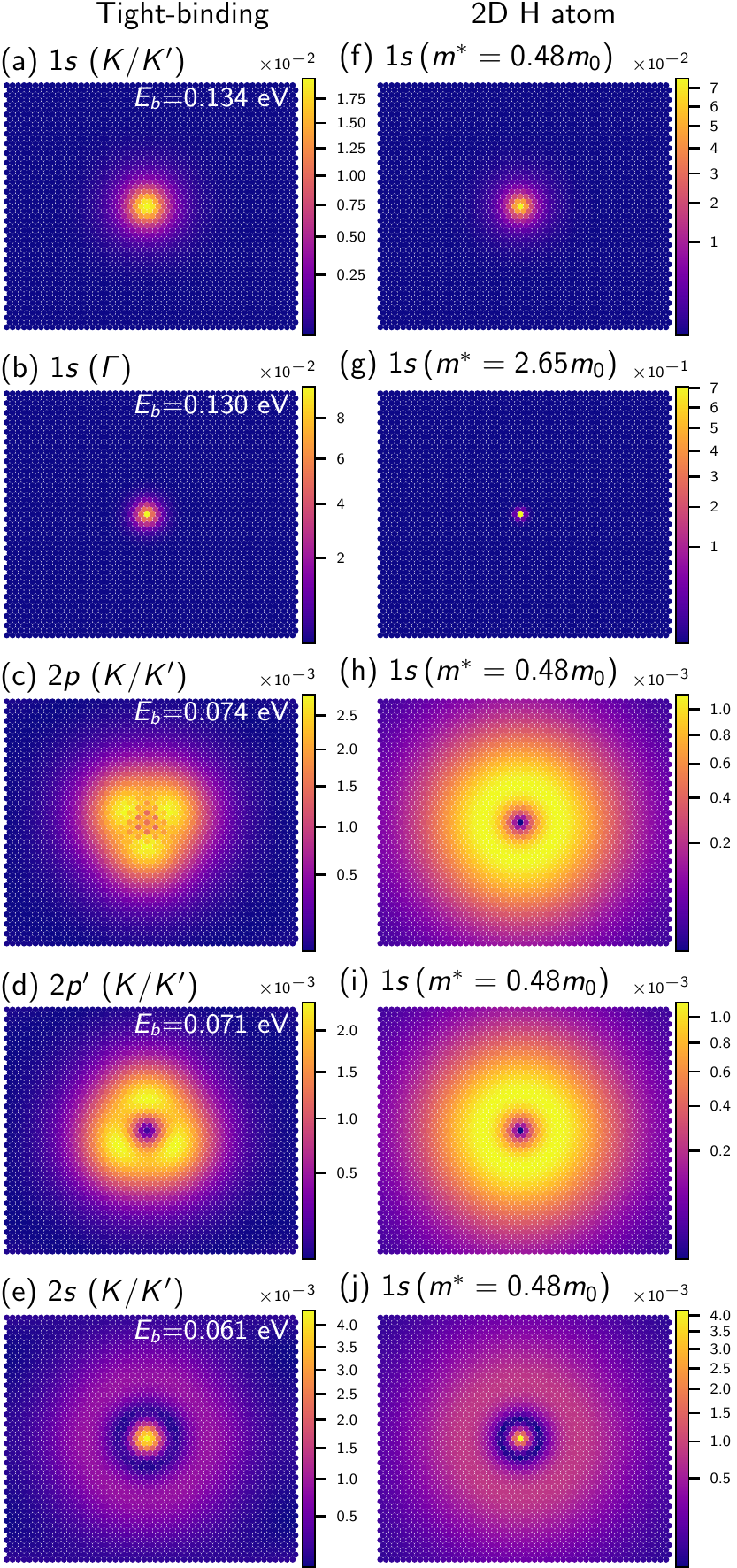}
\caption{(a)-(e) Squared wavefunctions of bound impurity states (TB model with RPA-screened potential), for an impurity charge $Q=-0.3e$ placed $2\text{ \AA}$ above the Mo site. States are labelled by their 2D hydrogenic character and origin in the BZ, found by projection onto the unperturbed states (see supplementary material). The corresponding binding energies $E_{\text{b}}$ with respect to the VBM are given in white. (f)-(j) 2D hydrogenic states with a nuclear charge of $Q=-0.3\zeta e$ (with $\zeta$ being the ratio of the screened and unscreened potentials at $r=0$ in Fig.~\ref{fig:eps}) for comparison, labelled by the effective mass of the VBM from which the corresponding states in (a)-(e) originate.}
\label{fig:imp1}
\end{figure}

\noindent\textbf{Acceptor States} - Figs.~\ref{fig:imp1}(a)-(e) show the wavefunctions (specifically, their squared magnitudes sampled at the $\Gamma$-point of the first Brillouin zone) of the five most strongly bound impurity states for an adatom with $Z=-0.3$, situated $d=2\text{ \AA }$ above the Mo-site, as calculated from our tight-binding model with an RPA-screened impurity potential. To label the impurity states, we compare them to the 2D hydrogenic states\cite{Yang1991}. While the two most strongly bound impurity states (Figs.~\ref{fig:imp1}(a) and (b)) have $1s$ character, the states in Figs.~\ref{fig:imp1} (c), (e) and (d) resemble the $2p$ and $2s$ states of the 2D hydrogen atom, respectively. We also present the corresponding 2D hydrogenic states in Fig.~\ref{fig:imp1}(f-j) for a nuclear charge $Q=-0.3\zeta$, where $\zeta\approx0.26$ is the ratio of the screened and unscreened potentials at $r=0$ in Fig.~\ref{fig:eps}. Surprisingly, the more strongly bound $1s$ states of Fig.~\ref{fig:imp1}(a) is significantly \emph{more delocalized} with an impurity radius of $a_\text{imp}=12.6$~\AA~than the less strongly bound $1s$ state in Fig.~\ref{fig:imp1}(b), which has a radius of $a_\text{imp}=5.19$~\AA. We determine $a_{\text{imp}}$ by fitting the impurity state to an exponential decay as in Eqn.~\ref{eqn:envelope}, and extracting the inverse decay scale $\alpha=a_{\text{imp}}^{-1}$. The $2p$ impurity states exhibit an angular modulation caused by the trigonal warping of the valence states near the band edge\cite{Kormanyos2013}. Note that the modulation is different for the two $2p$ states and we therefore label the second state distinctly as $2p^{\prime}$. In contrast to the 2D hydrogen atom, the $2s$, $2p$ and $2p^{\prime}$ are not degenerate, as indicated by their binding energies given in the top right corner of Fig.~\ref{fig:imp1}(a-e), because the impurity potential is screened and no longer follows a simple $1/r$ behaviour. 

\begin{figure}[ht!]
\centering
\includegraphics[]{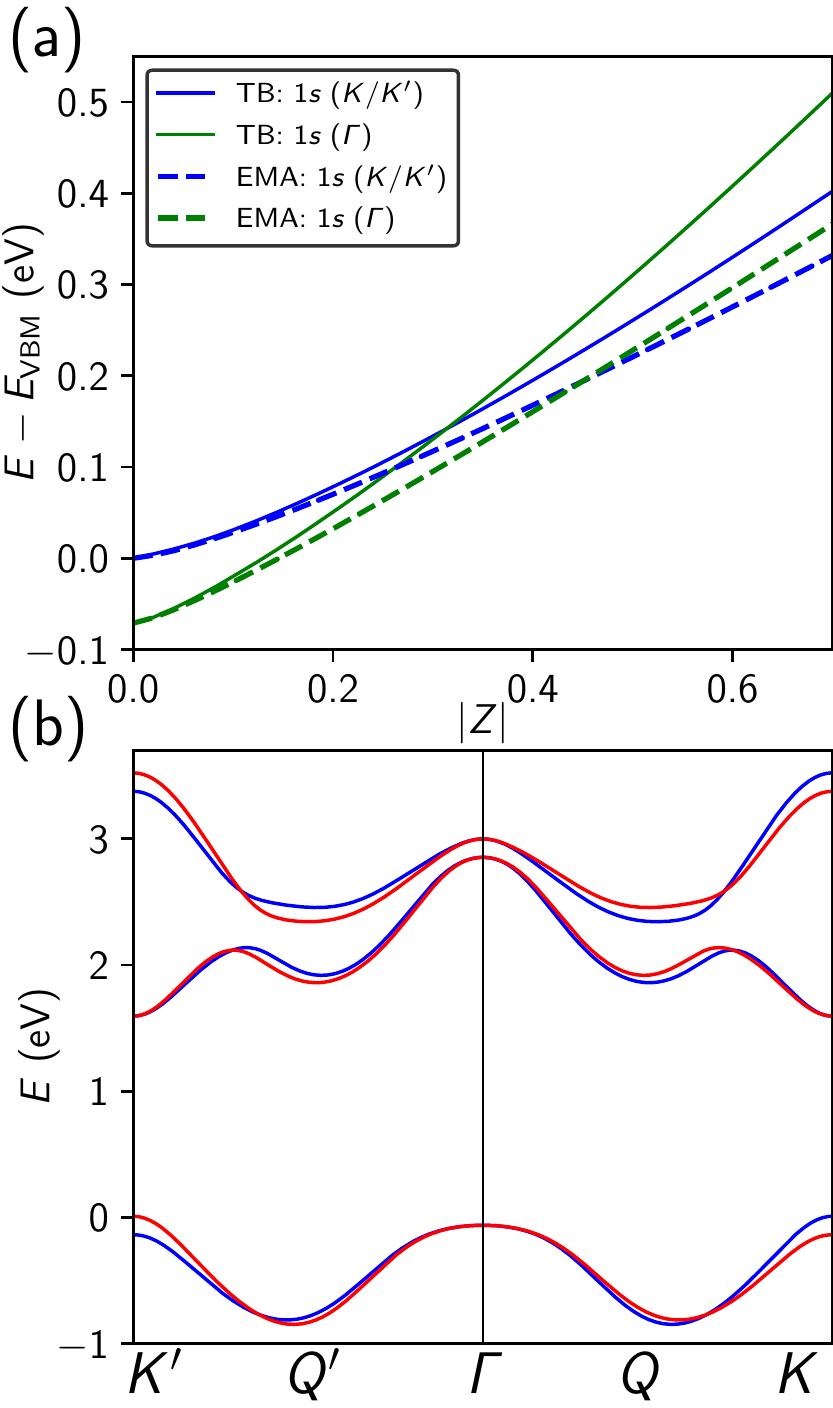}
\caption{(a) Binding energy $E_{\text{b}}=E-E_{\text{VBM}}$ of the $1s\;(K/K')$ (blue) and $1s\;(\Gamma)$ (green) impurity states as a function of adatom charge $Z$ for negatively charged adatoms on MoS$_2$ from tight-binding calculations (solid lines) and the effective mass approximation (EMA) (dashed lines). (b) Tight-binding band structure, where bands with spin-up (spin-down) character are in red (blue).}
\label{fig:ener1}
\end{figure}

To further analyze the impurity states, we projected their wavefunctions onto unperturbed states of the MoS$_2$ monolayer (see supplementary materials for details) and find that the most strongly bound $1s$ state and also the $2p$ and $2s$ states are composed of valence states from the $K$ and $K'$ points of the MoS$_2$ bandstructure, see Fig.~\ref{fig:ener1}(b). In contrast, the second $1s$ state originates from the valence band near the $\Gamma$-point of the unperturbed band structure. We label the states in Figs.~\ref{fig:imp1}(a-e) by their origin in the Brillouin zone (BZ), in addition to their 2D hydrogenic orbital character. We have subsequently labelled Figs.~\ref{fig:imp1}(f-j) by the effective mass of the valence band maxima (VBM) from which the corresponding TB states originate.

Fig.~\ref{fig:ener1}(a) shows the dependence of the impurity state binding energies $E_{\text{b}}=E-E_{\text{VBM}}$ (energy $E$ with reference to the primary valence band maximum $E_{\text{VBM}}$) on the adatom charge $Z$ for negatively charged adatoms. We have fitted the $1s$ binding energies to a power law of the form $-B+AZ^\eta$, see Table~\ref{tab:param}, where $B=0$ for $1s\;(K/K')$ and $B=0.071$~eV for $1s\;(\Gamma)$, and find that the $1s\;(K/K')$ and $1s\;(\Gamma)$ states have exponents of $\eta=1.30$ and $\eta=1.25$, respectively. These are significantly smaller than the exponent for a 2D hydrogen atom where the binding energy is given by $E(Z)=-4\frac{m^*}{m_0}Z^2$~Ry. Interestingly, the different Z-dependences of the $1s\;(K/K')$ and $1s\;(\Gamma)$ binding energies result in a \emph{crossover} at $Z=-0.32$, where the order of the two states switches. As the character of $1s\;(K/K')$ is dominated by Mo $4d_{xy}$ and $4d_{x^2-y^2}$ orbitals, while Mo $4d_{z^2}$ orbitals make up the $1s\;(\Gamma)$ state \cite{liu15}, our calculations suggest the possibility of controlling the orbital character of low-lying electronic states via defect engineering with potentially interesting consequences for optical properties.   

\begin{table}[ht!]
\centering
\caption{Coefficients of acceptor state binding energy fits given by $E_{\text{b}}=-B+AZ^\eta$ from tight-binding (TB) and effective mass theory (EMA) with the Keldysh model. All energies are referenced to the valence band maximum. We also show the impurity state radius $a_\text{imp}(Z)=\alpha^{-1}(Z)$  of the $1s$ states for $Z=-0.3$.}
\begin{tabular}{ c | c | c | c }
& $A$ (eV) & $\eta$ & $a_\text{imp}(-0.3)$ (\AA) \\
\hline
TB: $1s\;(K/K')$ & 0.641 & 1.30 & 12.6 \\
EMA: $1s\;(K/K')$  & 0.519 & 1.24 & 15.9 \\
TB: $1s\;(\Gamma)$  & 0.907 & 1.25 & 5.19 \\
EMA: $1s\;(\Gamma)$  & 0.661 & 1.15 & 6.65\\
\end{tabular}
\label{tab:param}
\end{table}

To further analyze the results of the tight-binding calculations, the bound impurity states were studied with effective mass theory. Specifically, we determined the impurity states associated with the subzones near $\Gamma$, $K$ and $K'$ using Eq.~\eqref{eqn:envelope}. For the acceptor states, each subzone acts as an independent 2D hydrogen-like system as the different spin states of the degenerate valence band maxima at $K$ and $K'$ prohibit interactions between the subzones. The resulting binding energies agree reasonably well with the tight-binding results, see dashed lines in Fig.~\ref{fig:ener1}(a) and Table~\ref{tab:param}. We see that the discrepancy between these two models increases with $Z$, as the RPA-screened potential in Fig.~\ref{fig:eps} is deeper than the screened potential in the Keldysh model, resulting in more strongly bound states. In particular, effective mass theory also predicts a crossover of $1s\;(K/K')$ and $1s\;(\Gamma)$ near $Z=-0.45$. The binding energy of $1s\;(\Gamma)$ increases more quickly with $Z$ because the effective mass near $\Gamma$ is about $5.5$ times larger than the effective mass near $K$ or $K'$. This also explains the differences in impurity radii, see Figs.~\ref{fig:imp1}(a) and (b).

\begin{figure}[h!]
\includegraphics[]{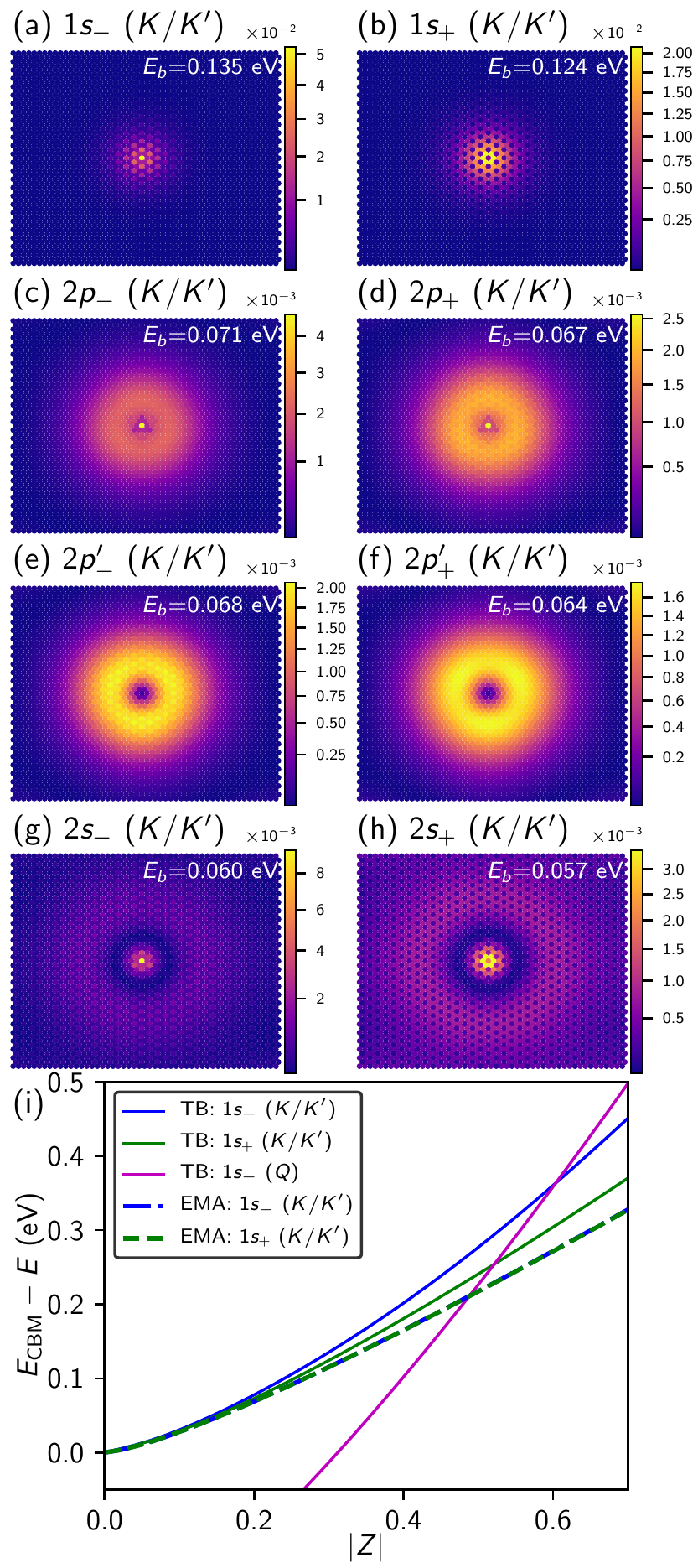}
\caption{(a-h) Squared wavefunctions of bound impurity states for an impurity charge $Q=+0.3e$ placed $2\text{ \AA}$ above the Mo site, with binding energies $E_{\text{b}}=E_{\text{CBM}}-E$ indicated (white). Hybridised states are separately labelled with $\pm$ subscripts. (i) Binding energy $E_{\text{b}}$ of hybridized $1s\;(K/K')$ (green and blue) and $1s\;(Q)$ (magenta) impurity states as a function of adatom charge $Z$ for positively charged adatoms on MoS$_2$ from TB (solid lines) and EMA (dashed lines).}
\label{fig:imp2}
\end{figure}

\noindent\textbf{Donor states} - Next, we study the shallow impurity states induced by positively charged adatoms. Figs.~\ref{fig:imp2}(a-h) show the wavefunctions of the eight most strongly bound impurity states for an adatom with $Z=0.3$ and $d=2\text{ \AA}$. The states are labelled based on their similarity to the eigenstates of the 2D hydrogen atom. In contrast to the acceptor case, we find \emph{a pair of states} corresponding to each solution of the 2D hydrogen atom, with different binding energies, indicated at the top right corner of each subfigure in white. The states of each pair are distinguished by a ``+'' or ``$-$'' subscript.

Fig.~\ref{fig:imp2}(i) shows the binding energies $E_{\text{b}}=E_{\text{CBM}}-E$ of the most strongly bound states (with energy $E$) with respect to the conduction band minimum (with energy $E_{\text{CBM}}$) as function of the impurity charge $Z$. At low values of $Z$, the $1s_-(K/K')$ and $1s_+(K/K')$ states are almost degenerate, but their binding energy difference increases with increasing $Z$.  A third impurity state originating from the local conduction band minimum at the 6 $Q$ points of the Brillouin zone crosses the two $1s\;(K/K')$ states near $Z=0.6$ and becomes the most strongly bound state for higher values of $Z$. The crossover is again caused by the larger effective mass at $Q$ point compared to the $K$ and $K'$ points. We have fitted the binding energies of these states to a power law of the form $B + AZ^\eta$, see Table~\ref{tab:param2}, where $B=0$ for states from $K/K'$ and $B=0.267$~eV for states from the $Q$-points. As for the acceptor impurity states, the exponents of the donor states are significantly smaller than the 2D hydrogen value $\eta=2$. 

\begin{table}[b!]
\centering
\caption{Coefficients of donor state binding energy fits given by $E_{\text{b}}=-B+AZ^\eta$ from tight-binding (TB) and effective mass theory (EMA) with the Keldysh model. All energies are referenced to the valence band maximum. We also show the impurity state radius $a_\text{imp}(Z)=\alpha^{-1}(Z)$ for $Z=0.3$.}
\begin{tabular}{ c | c | c | c }
& $A$ (eV) & $\eta$ & $a_\text{imp}(0.3)$ (\AA)\\
\hline
TB: $1s_-(K)$ &0.743 & 1.42 & 12.7 \\
EMA: $1s_-(K)$ & 0.513 & 1.24 & 15.4 \\
TB: $1s_+(K)$  & 0.588 & 1.29 & 16.4 \\
EMA: $1s_+(K)$ & 0.511 & 1.24 &  15.4\\
TB: $1s_-(Q)$ & 1.217 & 1.30 & ---\\
\end{tabular}
\label{tab:param2}
\end{table}
Again, we compare the tight-binding results to effective mass theory. We first determine the subzone envelope functions, Eq.~\eqref{eqn:envelope}, for the regions near the critical points at $K$ and $K'$. In contrast to the valence bands, there is no spin-orbit splitting of the conduction band states at $K$ and $K'$. As a consequence, the conduction band states at $K$ and $K'$ with equal spin are degenerate and this gives rise to the observed pairs of impurity states with same symmetry in Fig.~\ref{fig:imp2}. The subzone impurity states can couple and the resulting binding energy splitting is given by \cite{Bassani1974}
\begin{align}
\Delta_{KK'} \approx 2\left|\phi_{1s,\mathbf{K}}^*(r=0)\phi_{1s,\mathbf{K}'}(r=0) V(\mathbf{q}=\mathbf{K}-\mathbf{K}')\right|, 
\label{eqn:split}
\end{align}
We evaluate the splitting with the Keldysh approximation for $V$, using the Fourier transform of the screened Coulomb potential in the Keldysh model. We find that the splitting is several orders of magnitude smaller than the splitting found in the tight-binding model. This discrepancy is caused by the inaccurate behaviour of the Keldysh model at large wave vectors, which is shown in the inset of Fig.~\ref{fig:eps}, where the vertical black line indicates $|\mathbf{K}-\mathbf{K}'|$. We show the binding energies, found from effective mass theory using the Keldysh screening model for the splitting (see Fig.~\ref{fig:imp2}(i) as blue dashed an green dot-dashed lines). The fitting parameters of the binding energies to a power law are compared to the tight-binding results in  Table~\ref{tab:param2}. 

\begin{figure*}[ht!]
\centering
\includegraphics[]{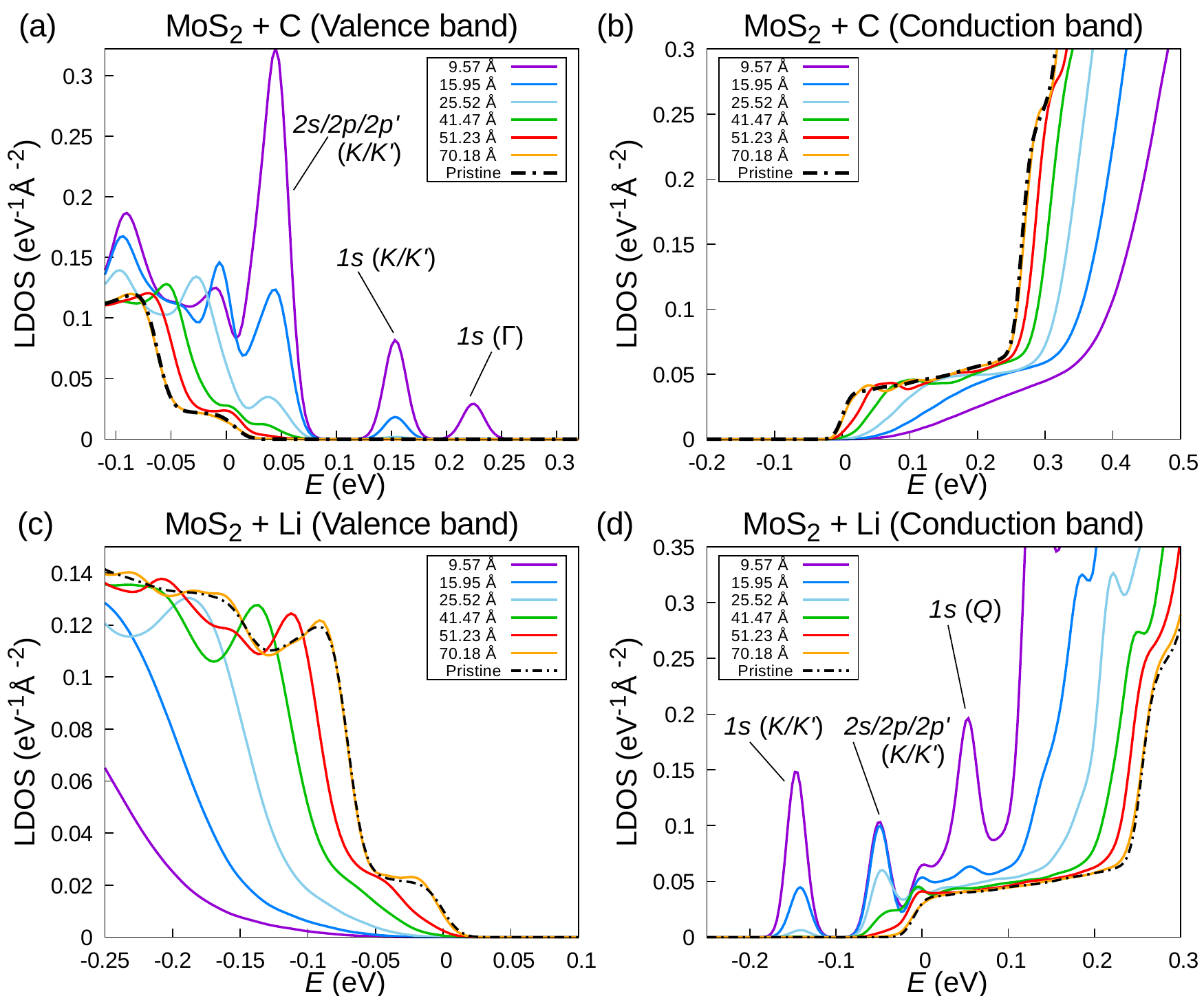}
\caption{(a)-(b): LDOS of a lithium (Li) adatom on MoS$_2$ ($+$ SiO$_2$ substrate) near the (a) valence band and (b) conduction band edge. (c)-(d): LDOS of a carbon (C) adatom on MoS$_2$ ($+$ SiO$_2$ substrate) near the (c) valence band and (d) conduction band edge. Results are shown for several distances from the impurity. In each graph, the zero of energy is set to the band edge of the unperturbed MoS$_2$.}
\label{fig:ldos}
\end{figure*}

The $1s$ impurity state wavefunctions from effective mass theory are given by 
\begin{align}
\Psi_{1s_\pm(K/K')}(\mathbf{r}) = \frac{1}{\sqrt{2}} \left( \phi_{1s,\mathbf{K}}(r)\psi_{\mathbf{K}}(\mathbf{r}) \pm \phi_{1s,\mathbf{K'}}(r)\psi_{\mathbf{K'}}(\mathbf{r}) \right),
\end{align}
where $\psi_{K/K'}(\mathbf{r})$ denote the Bloch states of the unperturbed MoS$_2$ band structure at $K$ and $K'$. Notably, the states with an $s$-character (Figs.~\ref{fig:imp2}(a), (b), (d) and (h)) exhibit an intensity modulation with a period of three unit cells along the directions connecting nearest neighbours. Projecting the impurity states onto unperturbed Bloch states reveals that all states originate from both the $K$ and $K'$ points of the Brillouin zone, where the minimum of the conduction band occurs, see Fig.~\ref{fig:ener1}(b). The corresponding probability densities contain a term with a $\cos((\mathbf{K}-\mathbf{K'})\cdot \mathbf{r})$ factor which gives rise to the oscillatory pattern in Figs.~\ref{fig:imp2}(a,e,d,h). In contrast to the impurity states with $s$-character which derive from unperturbed states directly at $K$ and $K'$, the states with $p$-character mostly derive from conduction band states in the vicinity of the band edges. As a consequence, the coupling between $K$ and $K'$ is weaker for the $p$-states and the spatial modulation is not observed. We find that this modulation does not occur when the defect is not placed on the transition-metal site. 

\noindent\textbf{Local density of states} - Scanning tunnelling spectroscopy (STS) provides spatially-resolved information about the electronic structure of surfaces and has been used to study the properties of shallow impurity states induced by charged adatoms experimentally. The $\mathrm{d}I/\mathrm{d}V$ curves obtained in STS are often assumed to be proportional to the local density of states (LDOS) of the sample. We have calculated the LDOS for values of $Z$ and $d$ that represent lithium (Li) and carbon (C) atoms adsorbed on a MoS$_2$. For Li, Chang \emph{et al.} found an impurity charge of $Z_{\text{Li}}=0.67$ from a Bader charge analysis\cite{Henkelman2006} of the DFT charge density \cite{chang14}. Using a similar procedure, Ataca \emph{et al.} determined $Z_{\text{C}}=-0.58$ for a C atom adsorbed to MoS$_2$ above the Mo site\cite{Ataca2011,He2010}. We modelled adsorbed atoms sitting above the Mo site at a height of $d_{\text{Li}}=3.1\text{ \AA}$ and $d_{\text{C}}=1.58\text{ \AA}$ \cite{chang14,Rastogi2014,Ataca2011,He2010}. Screening by a SiO$_2$ substrate is included via a substrate dielectric function of 3.7.   

Figs.~\ref{fig:ldos}(a-b) show the tight-binding LDOS for a C adatom on MoS$_2$ in the vicinity of valence band maximum and the conduction band minimum, respectively. A $6 \times 6$ $k$-point mesh and a Gaussian broadening of 0.01~eV were used. Near the VBM, several peaks originating from bound acceptor states can be observed in the band gap. The peak from $1s\;(\Gamma)$ disappears more quickly as a function of distance from the adatom than the $1s\;(K/K')$ peak. This is a consequence of the stronger localization of this state, see Fig.~\ref{fig:imp1}. At a distance of $\sim 66$~\AA~from the adatom, the LDOS of the perturbed system has converged to the LDOS of the pristine TMDC. In the vicinity of the CBM, no impurity states are present. However, the screened potential created by the adatom leads to a shift of the unperturbed LDOS.  

Fig.~\ref{fig:ldos}(c) and Fig.~\ref{fig:ldos}(d) show the tight-binding LDOS for a Li adatom on MoS$_2$ in the vicinity of valence band maximum and the conduction band minimum, respectively.  The peaks near the CBM in the vicinity of the adatom originate from bound donor states and can be observed up to a distance of $\sim 25$~\AA~from the adatom. Note that the splitting of the two impurity states from the $K$ and $K'$ points is too small to be resolved. No impurity state peaks are found in the vicinity of the VBM, but again the impurity potential causes a shift of the TMDC LDOS.

\textbf{Conclusions} - In summary, we have calculated properties of bound states induced by charged adatoms on monolayer MoS$_2$ using large-scale tight-binding simulations with screened impurity potentials from ab initio dielectric functions. We find that bound state wavefunctions exhibit symmetries similar to the eigenstates of the 2D hydrogen atom, but have radii of up to several nanometers because of electronic screening. Unconventional screening of the adatom charge also gives rise to significant deviations of the impurity state binding energies from the hydrogenic behaviour. In particular, we find that the dependence of the binding energies on the adatom charge $Z$ is described by a power law $Z^\eta$ with $\eta$ significantly smaller than two. Additional complexity arises from the multivalley band structure of MoS$_2$. For the acceptor states, a crossover occurs at a critical adatom charge where an impurity state from the $\Gamma$ valley becomes more strongly bound than states from the $K$ and $K'$ valleys. For the donor states, a similar crossover is observed between states from $K$, $K'$ and $Q$ valleys. These crossovers also lead to changes of the orbital character of the lowest impurity with potentially significant consequences for optical properties. Absence of spin-orbit interactions for conduction states at $K$ and $K'$ allows hybridization between donor impurity states from these valleys resulting in a $Z$-dependent splitting. We have compared our results to effective mass theory calculations with the Keldysh screening model and observe significant quantitative discrepancies, in particular for the splitting of hybridized donor states. We also present results for the local density of states for carbon and lithium adatoms which can be measured in scanning tunnelling experiments. Our calculations demonstrate the potential of adsorbate engineering for ultrathin devices based on TMDCs. 

\textbf{Acknowledgements} - This work was supported through a studentship in the Centre for Doctoral Training on Theory and Simulation of Materials at Imperial College London funded by the EPSRC (EP/L015579/1). We acknowledge the Thomas Young Centre under grant number TYC-101. This work used the ARCHER UK National Supercomputing Service (http://www.archer.ac.uk), and the Imperial College London High-Performance Computing Facility.

\bibliography{localized}

\begin{thebibliography}{10}

\bibitem{liu15}
G.~Liu, D.~Xiao, Y.~Yao, X.~Xu, and W.~Yao.
\newblock Electronic structures and theoretical modelling of two-dimensional
  group-vib transition metal dichalcogenides.
\newblock {\em Chem. Soc. Rev.}, 44:2643–--2663, 2015.

\bibitem{Kadantsev2012}
E.~S. Kadantsev and P.~Hawrylak.
\newblock Electronic structure of a single mos2 monolayer.
\newblock {\em Solid State Comm.}, 152(10):909--913, 2012.

\bibitem{Wang2012}
Q.~H. Wang, K.~Kalantar-Zadeh, A.~Kis, J.~N. Coleman, and M.~S. Strano.
\newblock Electronics and optoelectronics of two-dimensional transition metal
  dichalcogenides.
\newblock {\em Nature NanoTech.}, 7:699--712, 2012.

\bibitem{Xu2017}
J.~Xu, L.~Chen, Y.~Dai, Q.~Cao, Q.~Sun, and S.~Ding.
\newblock A two-dimensional semiconductor transistor with boosted gate control
  and sensing ability.
\newblock {\em Sci. Adv.}, 3(1602246):1--8, 2017.

\bibitem{Wachter2017}
S.~Wachter, D.~K. Polyushkin, O.~Bethge, and T.~Mueller.
\newblock A microprocessor based on a two-dimensional semiconductor.
\newblock {\em Nat. Comms.}, 8(14948):1--6, 2017.

\bibitem{Baugher2014}
B.~W.~H. Baugher, H.~O.~H. Churchill, Y.~Yang, and P.~Jarillo-Herrero.
\newblock Optoelectronic devices based on electrically tunable p–n diodes in
  a monolayer dichalcogenide.
\newblock {\em Nat. Nano.}, 9(April):262--267, 2014.

\bibitem{LopezSanchez2013}
O.~Lopez-Sanchez, D.~Lembke, M.~Kayci, A.~Radenovic, and A.~Kis.
\newblock Ultrasensitive photodetectors based on monolayer mos2.
\newblock {\em Nat. Nano.}, 8(July):497--501, 2013.

\bibitem{Kalantar2016}
K.~Kalantar-Zadeh and J.~Z. Ou.
\newblock Biosensors based on two-dimensional mos2.
\newblock {\em ACS Sens.}

\bibitem{Review2015}
Z.~Lin, B.~R. Carvalho, E.~Kahn, R.~Lv, R.~Rao, H.~Terrones, M.~A. Pimenta, and
  M.~Terrones.
\newblock {\em 2D Materials}, 3(022002), 2016.

\bibitem{kuc15}
A.~Kuc, T.~Heine, and A.~Kis.
\newblock Electronic properties of transition-metal dichalcogenides.
\newblock {\em MRS Bulletin}, 40:577--584, 2015.

\bibitem{SchmidtMende2007}
L.~Schmidt-Mende and J.~L. Macmanus-Driscoll.
\newblock Zno – nanostructures, defects, and devices.
\newblock {\em Materials Today}, 10(5):40--48, 2007.

\bibitem{Queisser1998}
H.~J. Queisser and E.~E. Haller.
\newblock Defects in semiconductors: Some fatal , some vital.
\newblock 281(5379):945--950, 1998.

\bibitem{Janotti2009}
A.~Janotti and C.~G. Van~de Walle.
\newblock Fundamentals of zinc oxide as a semiconductor.
\newblock {\em Rep. Prog. Phys.}, 72(126501), 2009.

\bibitem{Chin2012}
K.~K. Chin.
\newblock Dual roles of doping and trapping of semiconductor defect levels and
  their ramification to thin film photovoltaics.
\newblock {\em J. Appl. Phys.}, 111, 2012.

\bibitem{Leijtens2016}
T.~Leijtens, G.~E. Eperon, A.~J. Barker, G.~Grancini, W.~Zhang, J.~M. Ball,
  R.~Srimath, J.~Snaith, and A.~Petrozza.
\newblock Carrier trapping and recombination: the role of defect physics in
  enhancing the open circuit voltage of metal halide perovskite solar cells.
\newblock {\em Energy {\&} Environmental Science}, 9:3472--3481, 2016.

\bibitem{Komesu2017}
T.~Komesu, D.~Le, I.~Tanabe, E.~F. Schwier, Y.~Kojima, M.~Zheng, K.~Taguchi,
  K.~Miyamoto, T.~Okuda, H.~Iwasawa, K.~Shimada, T.~S. Rahman, and P.~A.
  Downben.
\newblock Adsorbate doping of mos2 and wse2: The influence of na and co.
\newblock {\em J. Phys.: Cond. Matt.}, 29(285501):0--7, 2017.

\bibitem{Rastogi2014}
Priyank Rastogi, Sanjay Kumar, Somnath Bhowmick, Amit Agarwal, and Yogesh~Singh
  Chauhan.
\newblock Ab-initio study of doping versus adsorption in monolayer mos 2.
\newblock {\em Conference: Emerging Electronics (ICEE) IEEE 2nd International},
  118:30309--30314, 2014.

\bibitem{chang14}
J.~Chang, S.~Larentis, E.~Tutus, L.~F. Register, and S.~K. Banerjee.
\newblock Atomistic simulation of the electronic states of adatoms in monolayer
  mos2.
\newblock {\em Appl. Phys. Lett.}, 104(141603), 2014.

\bibitem{Ataca2011}
C.~Ataca and S.~Ciraci.
\newblock Functionalization of single-layer mos2 honeycomb structures.
\newblock {\em J. Phys. Chem. C}, pages 13303--13311, 2011.

\bibitem{Fang2013}
H.~Fang, M.~Tosun, G.~Seol, T.~C. Chang, K.~Takei, J.~Guo, and A.~Javey.
\newblock Degenerate n‑doping of few-layer transition metal dichalcogenides
  by potassium.
\newblock {\em Nano Lett.}, 13:1991--1995, 2013.

\bibitem{Douloui2013}
K.~Dolui, I.~Rungger, C.~D. Pemmaraju, and S.~Sanvito.
\newblock Possible doping strategies for mos2 monolayers: An ab initio study.
\newblock {\em Phys. Rev. B}, 88(075420), 2013.

\bibitem{Wong2017}
D.~Wong, F.~Corsetti, Y.~Wang, V.~W. Brar, H.~Tsai, Q.~Wu, R.~K. Kawakami,
  A.~Zettl, A.~A. Mostofi, J.~Lischner, and M.~F. Crommie.
\newblock Spatially resolving density-dependent screening around a single
  charged atom in graphene.
\newblock {\em Phys. Rev. B}, 95(205419), 2017.

\bibitem{corsetti17}
F.~Corsetti, A.~A. Mostofi, and J.~Lischner.
\newblock First-principles multiscale modelling of charged adsorbates on doped
  graphene.
\newblock {\em 2D Mater.}, 4(025070), 2017.

\bibitem{Bassani1969}
F.~Bassani, G.~Iadonisi, and B.~Preziosi.
\newblock Band structure and impurity states.
\newblock {\em Phys. Rev.}, 186:735--746, 1969.

\bibitem{kohn1957}
W~Kohn.
\newblock Shallow impurity states in silicon and germanium.
\newblock {\em Solid State Physics}, 5:257 -- 320, 1957.

\bibitem{Rak2009}
Z.~Rak, S.~D. Mahanti, K.~C Mandal, and N.~C. Fernelius.
\newblock Electronic structure of substitutional defects and vacancies in gase.
\newblock {\em J. Phys. and Chem. of Solids}, 70(2):344--355, 2009.

\bibitem{Shimizu1965}
I.~Shimizu.
\newblock Physics of semiconductors.
\newblock {\em Phys. Lett.}, 15(297), 1965.

\bibitem{liu13}
G.~Liu, W.~Shan, Y.~Yao, W.~Yao, and D.~Xiao.
\newblock Three-band tight-binding model for monolayers of group-vib transition
  metal dichalcogenides.
\newblock {\em Phys. Rev. B}, 88:085433, 2013.

\bibitem{Qui2016}
D.~Y. Qui, de~Jornada F.~H., and S.~G. Louie.
\newblock Screening andmany-body effects in two-dimensional crystals:monolayer
  mos2.
\newblock 2016.

\bibitem{Ismail2006}
S.~Ismail-Beigi.
\newblock {\em Phys. Rev. B}, 73(233103), 2006.

\bibitem{BohmPines}
D.~Bohm and D.~Pines.
\newblock A collective description of electron interactions: Ii. collective vs
  individual particle aspects of the interactions.
\newblock {\em Phys. Rev.}, 85(338), 1952.

\bibitem{Giannozzi2009}
P.~Giannozzi, S.~Baroni, N.~Bonini, M.~Calandra, R.~Car, C.~Cavazzoni,
  D.~Ceresoli, G.~L. Chiarotti, M.~Cococcioni, I.~Dabo, A.~Dal~Corso,
  S.~de~Gironcoli, S.~Fabris, G.~Fratesi, R.~Gebauer, U.~Gerstmann,
  C.~Gougoussis, A.~Kokalj, M.~Lazzeri, L.~Martin-Samos, N.~Marzari, F.~Mauri,
  R.~Mazzarello, S.~Paolini, A.~Pasquarello, L.~Paulatto, C.~Sbraccia,
  S.~Scandolo, G.~Sclauzero, A.~P. Seitsonen, A.~Smogunov, P.~Umari, and R.~M.
  Wentzcovitch.
\newblock Quantum espresso: a modular and open-source software project for
  quantum simulations of materials.
\newblock {\em J. Phys.: Cond. Matt.}, 21(395502), 2009.

\bibitem{deslippe2012}
J.~Deslippe, G.~Samsonidze, D.~A. Strubbe, M.~Jain, M.~L. Cohen, and S.~G.
  Louie.
\newblock Berkeleygw: A massively parallel computer package for the calculation
  of the quasiparticle and optical properties of materials and nanostructures.
\newblock {\em Comp. Phys Comms.}, 183(6):1269 -- 1289, 2012.

\bibitem{Lanoo92}
M.~Lannoo.
\newblock The theory of impurity states in semiconductors.
\newblock {\em Physica Scripta}, T45:135--139, 1992.

\bibitem{LiXia2007}
S.~Li and J.~Xia.
\newblock Electronic states of a hydrogenic donor impurity in semiconductor
  nano-structures.
\newblock {\em Phys. Lett. A}, 366:120--123, 2007.

\bibitem{Bassani1974}
F.~Bassani, G.~Iadonisi, and B.~Preziosi.
\newblock Electronic impurity levels in semiconductors.
\newblock {\em Rep. Prog. Phys.}, 37(9):1009--1210, 1974.

\bibitem{Yang1991}
X~L.~Yang, S~H.~Guo, F~T.~Chan, K~W.~Wong, and Wai-Yim Ching.
\newblock Analytic solution of a two-dimensional hydrogen atom. i.
  nonrelativistic theory.
\newblock {\em Phys. Rev. A}, 43(3):1186--1196, 1991.

\bibitem{Keldysh79}
L.~V. Keldysh.
\newblock Coulomb interaction in thin semiconductor and semimetal films.
\newblock {\em JETP Lett.}, 29:658, 1979.

\bibitem{Berkelbach2013}
Hybertsen M.~S. Berkelbach, T. and D.~R. Reichman.
\newblock Theory of neutral and charged excitons in monolayer transition metal
  dichalcogenides.
\newblock {\em Phys. Rev. B}, 88, 2013.

\bibitem{Cudazzo2010}
P~Cudazzo, C~Attaccalite, I~V Tokatly, and A~Rubio.
\newblock Strong charge-transfer excitonic effects and the bose-einstein
  exciton condensate in graphane.
\newblock {\em Phys. Rev. Lett.}, 104(226804), 2010.

\bibitem{Kormanyos2013}
A.~Korm{\'{a}}nyos, V.~Z{\'{o}}lyomi, N.~D. Drummond, P.~Rakyta, G.~Burkard,
  and V.~I. Fal'Ko.
\newblock Monolayer mos2: Trigonal warping, the "gamma"-valley, and spin-orbit
  coupling effects.
\newblock {\em Phys. Rev. B}, 88(045416):1--8, 2013.

\bibitem{Henkelman2006}
G.~Henkelman.
\newblock A fast and robust algorithm for bader decomposition of charge
  density.
\newblock 36(3):354--360, 2006.

\bibitem{He2010}
Jiangang He, Kechen Wu, Rongjian Sa, Qiaohong Li, and Yongqin Wei.
\newblock Magnetic properties of nonmetal atoms absorbed mos2 monolayers.
\newblock {\em Applied Physics Letters}, 96(8):082504, 2010.

\bibitem{Hamann2013}
D~R Hamann.
\newblock Optimized norm-conserving vanderbilt pseudopotentials.
\newblock {\em Phys. Rev. B}, 88(085117):1--10, 2013.

\bibitem{Wehling2011}
T.~O. Wehling, E.~Şaşıoğlu, C.~Friedrich, A.~I. Lichtenstein, M.~I.
  Katsnelson, and S.~Blügel.
\newblock Strength of effective coulomb interactions in graphene and graphite.
\newblock {\em Phys. Rev. Lett.}, 106, 2011.

\end{thebibliography}

\newpage
\section{Supplementary Material}

\subsection{Ab initio Adatom Potential}
\begin{figure}[t!]
\centering
\includegraphics[width=0.5\textwidth]{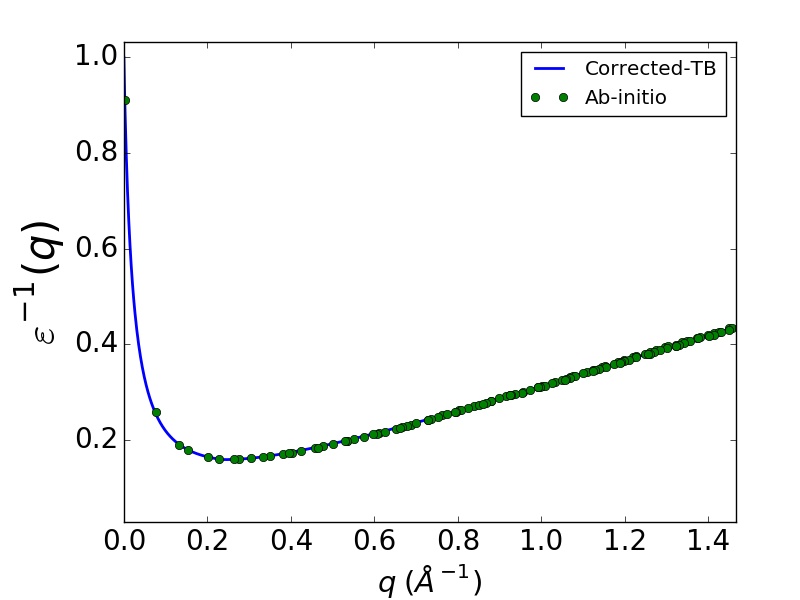}
\caption{Inverse dielectric function of MoS$_2$ for both a fitted functional form (blue curve) and sampled ab initio (green markers).}
\label{fig:fit}
\end{figure}

The screened potential of a charged adatom on a molybdenum disulfide (MoS$_2$) monolayer is obtained by first calculating the dielectric matrix $\varepsilon_{\mathbf{GG'}}(\mathbf{q})$ of an infinite stack of MoS$_2$ sheets and then calculating $\varepsilon^{-1}_{\text{2D}}(\mathbf{q})$ using 
\begin{equation}
\varepsilon^{-1}_{\text{2D}}(\mathbf{q}) = \frac{q}{2\pi e^2 L_z}\sum_{\mathbf{G}_z \mathbf{G}^{\prime}_z}\varepsilon^{-1}_{\mathbf{G}_z \mathbf{G}^{\prime}_z}(\mathbf{q})v_{\text{trunc}}(|\mathbf{q}+\mathbf{G}^{\prime}_z|),\label{eqn:eps2dq}
\end{equation}
where $\mathbf{G}_z$ and $\mathbf{G}^{\prime}_z$ denote reciprocal lattice vectors along the out-of-plane ($z$) direction, $v_{\text{trunc}}$ is a slab-truncated Coulomb interaction\cite{Ismail2006} and $L_z$ denotes the distance between the stacked sheets. To do this, we first perform density-functional theory (DFT) calculations within the generalized gradient approximation (GGA), using the Perdew-Burke-Ernzerhof (PBE) exchange-correlation functional and optimized norm-conserving Vanderbilt pseudopotentials\cite{Hamann2013}. Calculations were carried out using the Quantum Espresso software package\cite{Giannozzi2009}. To determine the ground state electron density, we use a $14 \times 14$ $\Gamma$-centred $k$-point mesh and an 80 Ry plane-wave cutoff. The stacked MoS$_2$ sheets are separated by $L_z=15.95\text{ \AA}$ in the out-of-plane direction. Next, the inverse dielectric matrix $\varepsilon^{-1}_{\mathbf{GG'}}(\mathbf{q})$ is calculated using the BerkeleyGW software package\cite{deslippe2012}, on a $30 \times 30$ $q$-point mesh using a plane-wave cut-off of 30 Ry and we sum over 2587 unoccupied states. The sampled points are shown in Fig.~\ref{fig:fit} as green circles, showing that the inverse dielectric function is isotropic at small wavevectors.

Having determined $\varepsilon^{-1}_{\text{2D}}(\mathbf{q})=\varepsilon^{-1}_{\text{2D}}(q)$ from first principles on a discrete $q$-point mesh, we fit the high-$q$ and low-$q$ regions to a functional form, and express the remaining mid-$q$ region as the sum of the inverse dielectric function of the tight-binding calculation and a correction in order to perform integrations in reciprocal space more easily. For $q<q_{\text{low}}$ with $q_{\text{low}}\approx 0.23\text{ \AA}^{-1}$, the sampled dielectric function is fitted to tanh$(x)$, as this has been previously used to describe the long-range screening of thin-film semiconductors\cite{Keldysh79}. This takes the form:
\begin{equation}
\varepsilon_{\text{low}}(q) = \kappa_1\,\text{tanh}\left(\frac{qh_1}{2}+\frac{1}{2}\text{ln}\left|\frac{\kappa_1+1}{\kappa_1-1}\right|\right),
\end{equation}
where we find $\kappa_1=6.68$ and $h_1=13.17\text{ \AA}$. For $q_{\text{low}}\leq q < q_{\text{high}}$, where $q_{\text{high}}=1.5\text{ \AA}^{-1}$, we represent $\varepsilon^{-1}_{\text{2D}}(\mathbf{q})$ using the  dielectric function of the three-band tight-binding model, given by 
\begin{equation}\label{eqn:adler}
\varepsilon_{\text{TB}}(\mathbf{q}) = 1 - \frac{v_{\mathbf{q}}}{\Omega}\sum_{n n'}\sum_{\mathbf{k}\in\text{BZ}}\frac{(f_{n\mathbf{k}}-f_{n'\mathbf{k+q}})|M_{nn'}(\mathbf{k},\mathbf{q})|^2}{E_{n\mathbf{k}}-E_{n'\mathbf{k+q}}},
\end{equation}
where $v_{\mathbf{q}}=2\pi/q$ is 2D Fourier transform of the Coulomb potential, $\Omega$ is the unit cell area, and $M_{nn'}(\mathbf{k},\mathbf{q})=\langle{\psi_{n\mathbf{k}}}|e^{-i\mathbf{q}\cdot\mathbf{r}}|\psi_{n'\mathbf{k+q}}\rangle$ is the matrix element. For $q<1.5\text{ \AA}^{-1}$, we find that $\varepsilon_\text{TB}^{-1}(\mathbf{q})\equiv 1/\varepsilon_\text{TB}(\mathbf{q})$ is highly isotropic and we carry out an angular average to obtain $\bar{\varepsilon}^{-1}_{\text{TB}}(q)$. To include the effect of the other bands on screening, we employ the correction proposed by Wehling \emph{at al.} \cite{Wehling2011} which captures to electrostatic screening of a thin film with thickness $h_2$ and dielectric constant $\kappa_2$ at long wavelengths: 
\begin{equation}
\delta\varepsilon^{-1}(q)=\frac{1}{\kappa_2}\frac{\kappa_2+1 + (\kappa_2-1)\,e^{-qh_2}}{\kappa_2+1 - (\kappa_2-1)\,e^{-qh_2}}-1.
\end{equation} 
The parameters $\kappa_2=0.69$ and $h_2=0.73\text{ \AA}$ were fitted to the ab initio inverse dielectric function. At large values of $q\geq q_{\text{high}}$, the tail of the dielectric function is fitted to $\varepsilon_{\text{high}}(q)=1+\chi_c/q$, where $\chi_c=1.83\text{ \AA}^{-1}$. In summary, we fit $\varepsilon^{-1}_{\text{2D}}(q)$ in Eq. \ref{eqn:eps2d} to the functional form:
\begin{equation*}
\varepsilon^{-1}_{\text{2D}}(q) = \begin{cases} & \varepsilon^{-1}_{\text{low}}(q),\;\;\;\;q <q_{\text{low}}, \\
& \bar{\varepsilon}_{\text{TB}}^{-1}(q) + \delta\varepsilon^{-1}(q; h, \kappa),\;\;\;\; q_{\text{low}}\leq q < q_{\text{high}},\\
& \varepsilon^{-1}_{\text{high}}(q),\;\;\;\; q\geq q_{\text{high}},
\end{cases}.
\end{equation*}
ensuring continuity between intervals. The corrected dielectric function $\varepsilon^{-1}_{\text{2D}}(q)$ is shown in Fig.~\ref{fig:fit}, comparing ab initio results with the corrected dielectric function. 

\subsection{Reciprocal-space Impurity Envelopes}
\begin{figure*}[h!]
\includegraphics[width=1.0\textwidth]{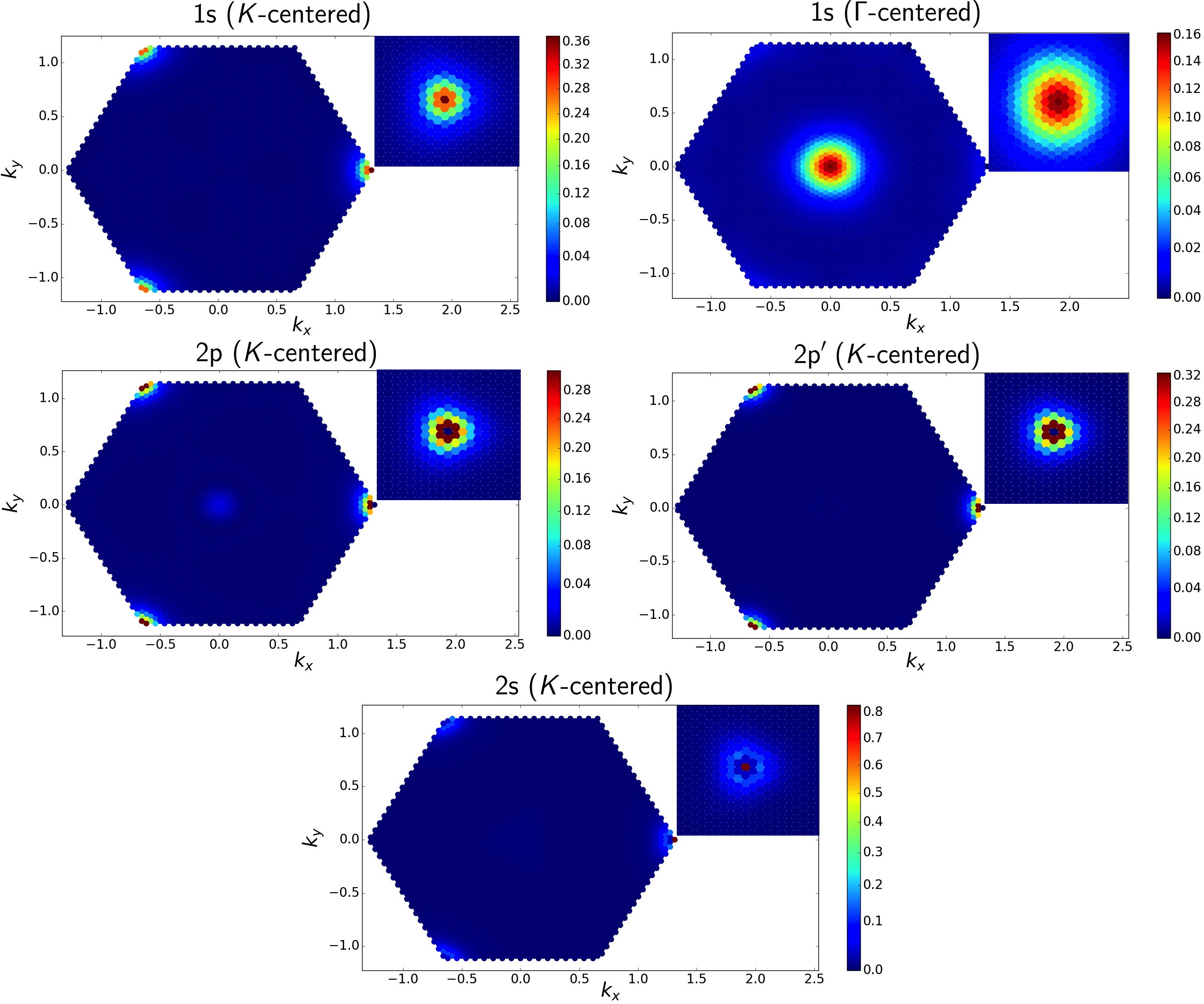}
\caption{Projections of top five impurity states in the Brillouin Zone, for a $Z=-0.3$ acceptor charge placed $d=2\text{ \AA}$ above a MoS$_2$ monolayer. The inset shows the projections centred on their respective high-symmetry points.}
\label{fig:proj1}
\end{figure*}

\begin{figure*}[h!]
\includegraphics[width=1.0\textwidth]{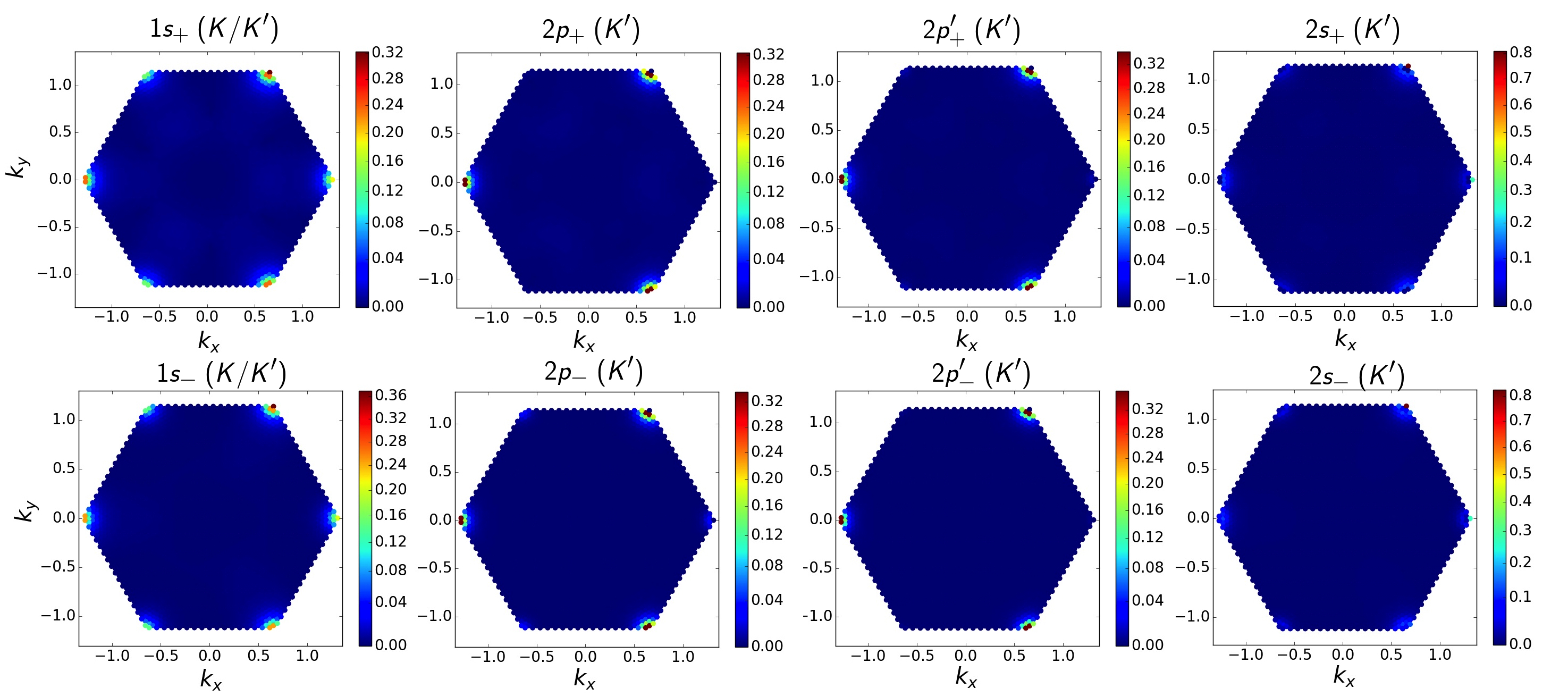}
\caption{Projections of top eight impurity states in the Brillouin Zone, for a $Z=0.3$ donor charge placed $d=2\text{ \AA}$ above a MoS$_2$ monolayer.}
\label{fig:proj2}
\end{figure*}

To identify the origin in the Brillouin zone (BZ) of impurity states, we construct unperturbed $N\times N$ supercell eigenstates using unit cell states $|\Psi^{\text{UC}}_{\mathbf{g}m}\rangle$ which fold onto the $\Gamma$ point, such that $\mathbf{g}=\mathbf{G}/N$ lies in the first BZ of the unit cell system and $m$ is the band index. For each $\mathbf{g}$, we create the set of eigenstates
\begin{equation*}
|\Psi^{(N)}_{\Gamma_ n}\rangle \equiv |\Psi^{(N)}_{\mathbf{g}m}\rangle = \frac{1}{N}\begin{pmatrix}
e^{i\mathbf{g}\cdot\mathbf{\tau}_0} \\
e^{i\mathbf{g}\cdot\mathbf{\tau}_1} \\
\vdots \\
e^{i\mathbf{g}\cdot\mathbf{\tau}_{N-1}}
\end{pmatrix}\bigotimes|\Psi^{\text{UC}}_{\mathbf{g}m}\rangle, 
\end{equation*}
where $\tau_i$ is position of the $i$th unit cell in the supercell, and $n$ now orders the folded eigenstates at $\Gamma$ in energy. We project the eigenstates of the perturbed supercell onto the set $\left\lbrace|\Psi^{(N)}_{\mathbf{g}m}\rangle\right\rbrace$ to determine the origin $\mathbf{g}$ in the BZ of the impurity states. For acceptor states at $Z=-0.3$, in Fig.~\ref{fig:proj1} we show the projections in the BZ. These are centered on their respective origins in the BZ, and demonstrate interesting localisation. For the $1s$ states, we see that the $\Gamma$ state is more delocalized in $k-$space than the state from $K$. The $2s$ states, originating  from $K$, demonstrate three-fold anisotropy attributed to the trigonal warping of the valence bands at $K$. This clear anisotropy manifests itself in the three-fold symmetric impurity state wavefunctions in Fig.~\ref{fig:imp1}, with orientation determined by the correspondence between the crystal lattice vectors and reciprocal lattice vectors. While Fig.~\ref{fig:proj1} shows only the absolute square of the projection for the $2p$ and $2p'$, the phases for the $2p$ and $2p'$ are opposite in sign and similar in value, resulting in a $\pi$ rotation of the $2p$ state onto the $2p'$ state. Anisotropy in the reciprocal-space impurity envelope occurs most prominently at inverse scales $\approx 0.06\text{ \AA}^{-1}-0.16\text{ \AA}^{-1}$, corresponding to trigonal lobes at $\approx 20 \text{ \AA}$ from the impurity bonding site. This anisotropy is not as prominent in the $2s$ state and negligible in the $1s$ state, which results in more isotropically distributed impurity wavefunctions. We have not shown the states at $K'$, as they contain the same information as Fig.~\ref{fig:proj2}. When the $K$ and $K'$ fold onto $\Gamma$, states from both points have a set of $1s, 2s, 2p/p'$ states with opposite spin in the three-band model. We also show the projection of the donor states for a charge of $Z=0.3$.

\end{document}